\documentstyle[12pt,epsfig]{article}
\newcommand{\be}[1]{\begin{equation} \label{(#1)}}
\newcommand{\ee}{\end{equation}}
\newcommand{\ba}[1]{\begin{eqnarray} \label{(#1)}}
\newcommand{\ea}{\end{eqnarray}}
\newcommand{\nn}{\nonumber}

\def\pmb#1{\setbox0=\hbox{#1}  \kern-.015em\copy0\kern-\wd0
  \kern.03em\copy0\kern-\wd0
  \kern-.015em\raise.0233em\box0 }

\def\rp{$R_p\hspace{-1em}/\ \ $}
\def\r{$R_p\hspace{-1em}/\hspace{3mm} $}

\def\Lfv{$L_f\hspace{-0.95em}/\ \ $}
\def\rpm{R_p \hspace{-0.8em}/\;\:}

\def\lg{\langle}
\def\rg{\rangle}

\def\m{$\mu^--e^-$}

\begin{document}
\hfill{USM-TH-111}\\[1cm] 
\begin{center}
{\Large\bf B-quark mediated neutrinoless $\mu^--e^-$ conversion in presence of R-parity violation}
\\

\bigskip

{T.S. Kosmas$^a$, Sergey Kovalenko$^{b}$\footnote{On
leave of absence from the Joint Institute for Nuclear Research, Dubna, Russia}
and Ivan Schmidt$^b$} \\[0.5cm]
{$^a$\it Division of Theoretical Physics, University of Ioannina GR-45110
Ioannina, Greece}\\[3mm]
{$^b$\it Departamento de F\'\i sica, Universidad
T\'ecnica Federico Santa Mar\'\i a, Casilla 110-V, Valpara\'\i so, Chile}

\end{center}

\bigskip

\begin{abstract}
We found that in supersymmetric models with R-parity non-conservation (\rp SUSY) 
the b-quarks may appreciably contribute to exotic neutrinoless  \m conversion in 
nuclei via the triangle diagram with two external gluons.
This allowed us to extract previously overlooked constraints on the third generation 
trilinear \r parameters significantly more stringent than those existing in the literature. 
\end{abstract}

\bigskip
\bigskip

PACS: 12.60Jv, 11.30.Er, 11.30.Fs, 23.40.Bw

\bigskip
\bigskip

KEYWORDS: Lepton flavor violation, $\mu -e$ conversion in nuclei,
supersymmetry, R-parity violation. 

\bigskip 
\bigskip

In the standard model (SM) the lepton flavors ($L_f$) are conserved quantum 
numbers as an accidental consequence of gauge invariance and field 
content. Thus, observation of $L_f$ non-conservation would imply the presence 
of physics beyond the standard model. Non-conservation of muon lepton flavor 
$L_{\mu}$ in neutrino oscillations has been recently established by 
the SuperKamiokande experiment in atmospheric neutrino measurements. 
In this case lepton flavor violation (\Lfv) is generated by non-zero neutrino masses.
Various sources of \Lfv can be probed by searching for certain exotic processes. 
Among them the neutrinoless muon-to-electron conversion in muonic atoms, 
$ \mu^- + (A,Z) \longrightarrow  e^- \,+\,(A,Z)^*$, is known to be one of the most powerful
tools to constrain \Lfv interactions 
\cite{KLV94}-\cite{KKS:2001}. 
In particular, it allows setting stringent constraints on the \Lfv 
interactions of the supersymmetric  models with R-parity violation (\rp SUSY). 
In the literature there have been obtained upper bounds on 
various products of the \Lfv  trilinear R-parity violating (\rp) couplings 
$\lambda\lambda'$, $\lambda'\lambda'$ \cite{9701381,Huitu,FKKV:2000,Barb}.
In the present letter we derive new constraints on the products
$\lambda\lambda'$ with some other combinations of generation indexes.
These constraints emerge from the previously overlooked contribution of b-quark 
to \m conversion via the triangle diagram shown in Fig. 1. 

 \begin{figure}[t!]
 \hspace{3.5cm}
  \mbox{\epsfxsize=8 cm\epsffile{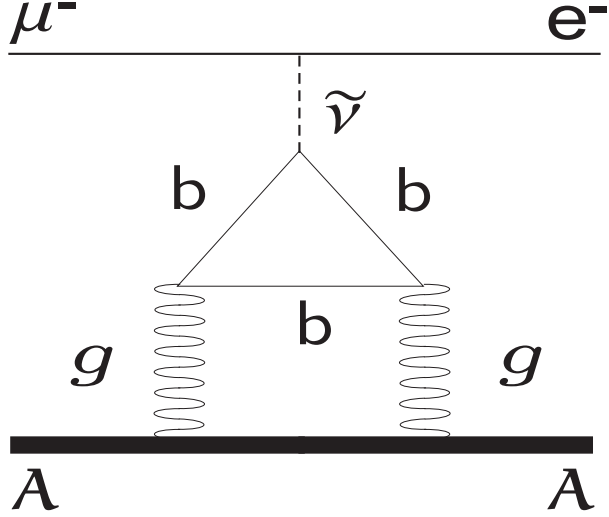}}

 \caption{The b-quark contribution to the nuclear \m conversion in the \rp SUSY.}
 \label{fig1}
 \end{figure} 

During the last decade \rp SUSY models have been extensively studied in the literature.
For the minimal field content the R-parity violating part of the superpotential reads
\begin{eqnarray}
W_{\rpm} =\lambda _{ijk}L_{i}L_{j}E_{k}^{c} + \bar\lambda _{ijk}^{\prime
}L_{i}Q_{j}D_{k}^{c}+\mu _{j}L_{j}H_{2}+\bar\lambda _{ijk}^{\prime \prime
}U_{i}^{c}D_{j}^{c}D_{k}^{c}.
\label{sup1}
\end{eqnarray}
The definition of the couplings $\bar\lambda', \bar\lambda'' $ corresponds to the gauge 
basis for the quark fields. We set $\bar\lambda^{\prime \prime }=0$, since these are 
irrelevant for our consideration. This ``ad hoc" condition ensures proton stability and 
can be guaranteed by special discreet symmetries other than R-parity. 

The leading quark-level tree diagrams with the trilinear \rp-couplings contributing 
to the $\mu-e$ conversion are listed in Ref. 
\cite{FKKV:2000} 
and are of the following three types:

(i) $\mu_{_{L,R}}\rightarrow e_{_{L,R}}, d_{_L}\rightarrow d_{_R}$, with t-channel $\tilde\nu$
exchange, 

(ii) $\mu_{_L}\rightarrow d_{_R}, d_{_R}\rightarrow e_{_L}$, with t-channel $\tilde{u}$
exchange,  

(iii) $\mu_{_L} u_{_L}\rightarrow \tilde{d}_{_R}\rightarrow e_{_L} u_{_L}$.

\noindent
The photonic  1-loop diagrams can also significantly contribute to this process \cite{Huitu}. 
However they are irrelevant for the present case of heavy quark contribution and, therefore, 
are not included in our analysis. Integrating out the heavy intermediate SUSY-particles from 
the above mentioned diagrams and carrying out a Fierz rearrangement  one obtains the following 
4-fermion effective Lagrangian for $\mu-e$ conversion at the quark level \cite{FKKV:2000}
\begin{equation}
{\cal L}_{eff}^{q}\ =\ \frac{G_F}{\sqrt{2}}\ j_{\mu }\left[ \eta^{ui}J_{u(i)}^{\mu }+
\eta^{di}J_{d(i)}^{\mu }\right] + 
\frac{G_F}{\sqrt{2}}\  (\eta_{L}^{di}\ j_L  + \eta_{R}^{di} j_R) \ J_{d(i)}.
\label{eff-q}
\end{equation}
The index $i$ denotes generation so that $u_i = u,c,t$ and $d_i = d,s,b$.
Here 
$J_{q(i)}^{\mu }=\bar{q}_i \gamma^{\mu } q_{i}$, 
$J_{d(i)}=\bar{d}_i d_{i} $, 
$j^{\mu }=\bar{e}\gamma^{\mu} P_L \mu$,
$j_{L,R} = \bar{e}P_{L,R} \mu$.
In Eq. (\ref{eff-q}) we neglected the terms with axial-vector and pseudoscalar 
quark currents which do not contribute to the dominant coherent mode of 
\m conversion \cite{KKS:2001,Chiang,Kosmas,tsk-tok}. 
The coefficients $\eta $ in Eq. (\ref{eff-q}) accumulate the dependence on 
$R_{p}\hspace{-1em}/\ \ $
SUSY parameters as 
\begin{eqnarray} \nn
\eta^{ui} &=&- \frac{1}{\sqrt{2}} \sum_{l,m,n}\frac{\lambda _{2ln}^{\prime}
\lambda_{1mn}^{\prime *}}{G_F \tilde m_{dR(n)}^{2}}V^*_{il} V_{im},  \ \  \ 
\eta^{di} =  \frac{1}{\sqrt{2}}\sum_{l,m,n}
\frac{\lambda_{2mi}^{\prime }\lambda_{1li}^{\prime *}}
{G_F \tilde m_{{u}L(n)}^{2}} V^*_{nm} V_{nl},  \\ 
\label{coeff12} 
\eta_{R}^{di} &=& - \sqrt{2} \sum_{n}
\frac{\lambda_{nii}^{\prime }\lambda _{n12}^{*}}{{G_F \tilde m^2_{\nu(n)}}}, 
\ \ \ \ \ \ \ \ \ \ \ \ \ \ \ 
\eta_{L}^{di} = - \sqrt{2} \sum_{n}
\frac{\lambda_{nii}^{\prime*}\lambda_{n21}}{G_F \tilde m^2_{{\nu}(n)}}.
\end{eqnarray}
Here $\tilde m_{q(n)}, \tilde m_{\nu(n)}$ are the squark and sneutrino
masses.
In Eq. (\ref{coeff12}) we introduced the couplings 
$\lambda_{ijk}^{\prime} = \bar\lambda_{imn}^{\prime }
\left(V^d_{_L}\right)^*_{jm} \left(V^d_{_R}\right)_{kn}$ corresponding to
the \r interactions in the quark mass eigenstate basis, 
related to the gauge basis $q'$ through 
$q_{_{L,R}} = V_{_{L,R}}^q q'_{_{L,R}}$. The CKM matrix
is defined in the standard way as  $V = V^u_{_L} V^{d\dagger}_{_L}.$

The contribution of the quark currents $J_{u(i)}^{\mu }, J_{d(i)}^{\mu }, J_{d(i)}$ present in Eq. (\ref{eff-q}) to the corresponding 
nucleon currents can be parametrized in the form
\begin{eqnarray}\label{q-nucl}
\langle N|\bar{q}\ \Gamma_{K}\ q|N\rangle = G_{K}^{(q,N)}
\bar{\Psi}_N\ \Gamma_{K}\ \Psi_N,
\end{eqnarray}
with $q=\{u,d,s\}$,  $N=\{p,n\}$ and  $K = \{V,S\}$,
$\Gamma_K = \{ \gamma_{\mu}, 1 \}$.
The maximum momentum transfer ${\bf q}$ in $\mu -e$ conversion can be estimated as  
$|{\bf q}| \approx m_\mu/c$ with $m_\mu=105.6$ MeV being the muon mass. Since $|{\bf q}|$ is 
relatively small compared to the typical nucleon structure scales 
we can safely neglect in Eq. (\ref{q-nucl})  the ${\bf q}^{2}$-dependence of 
the nucleon form factors $G_{K}^{(q,N)}$ as well as the weak magnetism and 
the induced pseudoscalar terms which are proportional to the small momentum transfer.

Isospin symmetry requires that 
$G_{K}^{(u,n)}=G_{K}^{(d,p)}\equiv G_{K}^{d}$,
$G_{K}^{(d,n)}=G_{K}^{(u,p)}\equiv G_{K}^{u}$, 
$G_{K}^{(s,n)}=G_{K}^{(s,p)}\equiv G_{K}^{s}$, 
$G_{K}^{(h,n)}=G_{K}^{(h,p)}\equiv G_{K}^{h}$, 
with $K=V,S$ and $h = c,b,t$. 
Furthermore, conservation of vector current requires the vector charge to be equal to 
the quark number of the nucleon. This allows fixing the vector nucleon constants as
$
G_{V}^{u}=2$, $G_{V}^{d}=1$, $G_{V}^{s}=G_{V}^{h}=0.
$
Thus the strange and heavy quarks can contribute only to the scalar nucleon current
\footnote{They can also contribute to the axial-vector and pseudoscalar nucleon currents which
are irrelevant for the coherent \m conversion considered in the present letter.}.
Since the scalar currents in Eq. (\ref{eff-q}) involve only down quarks $d,s,b$ it follows
that among the heavy $c,b,t$-quarks only the b-quark can contribute to the coherent \m conversion.  
The heavy quarks contribution to the scalar current is realized via the triangle diagram 
in Fig. 1 with the two gluon lines. The heavy quark $q_h$ scalar current induced by the diagrams 
of this type can be estimated using the heavy quark expansion  \cite{SVZ}
\begin{eqnarray}\label{expans}
m_h \ \bar{q}_h \ q_h \approx -\frac{2}{3} \left(\frac{\alpha_s}{8 \pi} \right) GG + 
O\left(\frac{\mu^2}{m_h^2} \right).
\end{eqnarray}
Here $\alpha_s$ and $\mu$ are the QCD coupling constant and a typical hadronic scale respectively,
and  
$G G  =  G_{\mu\nu}^a G^{\mu\nu}_a$ where 
$G_{\mu\nu}^a$ is the gluon field strength.
The quark scalar currents and the gluon operator $GG$ also contribute to 
the trace of the energy-momentum tensor
\begin{eqnarray}\label{trace}
\theta^{\mu}_{\mu} = m_u \bar u u + m_d \bar d d
+ m_s \bar s s  + \sum_h m_h \bar{q}_h q_h  - (b \alpha_s/8\pi)G G,
\end{eqnarray}
where $b = 11 - (2/3)n_f$ and $n_f=6$ is the number of quark species.

The scalar form factors $G_S^q$ can be extracted from the baryon octet $B$ mass
spectrum $M_B$, expressed as \cite{Cheng}
\begin{eqnarray}\label{emt}
\lg B|\theta^{\mu}_{\mu}|B\rg = M_B \bar B B,
\end{eqnarray}
and from the data on the pion-nucleon sigma term 
$
\sigma_{\pi N} = (1/2)(m_u + m_d)\lg p|\bar u u + \bar d d|p\rg .
$
Using Eqs. (\ref{expans})-(\ref{emt}) in combination with $SU(3)$ relations \cite{Cheng} for
the matrix elements in (\ref{emt}) we obtain 
\begin{eqnarray}\label{scalar}
G_S^{u} \approx 5.1, \ \ \ G_S^{d} \approx 4.3, \ \ \  G_S^{s} \approx 2.5, 
\ \ \ G_S^{b} \approx 9\times 10^{-3}.
\end{eqnarray}
Here we used for $\sigma_{\pi N} = 48$ MeV \cite{PiN}, and 
$m_u= 4.2$ MeV, $m_d = 7.5$ MeV, $m_s = 150$ MeV, $m_b = 4.2$ GeV.

Having the couplings $G_V^{q}, G_S^{q}$ determined
we rewrite the Lagrangian (\ref{eff-q}) in terms of the nucleon currents
\begin{equation}
{\cal L}_{eff}^{N}= \frac{G_F}{\sqrt{2}}\left[\bar{e}
\gamma_{\mu }(1-\gamma _{5})\mu \cdot J^{\mu }+
\bar{e}\mu \cdot J^{+}+\bar{e}\gamma_{5}\mu \cdot J^{-}\right].  \label{nucl1}
\end{equation}
Here we defined  
$
J^{\mu } = \bar{N}\gamma ^{\mu } (\alpha _{V}^{(0)}+\alpha_{V}^{(3)}\tau _{3}) \,N $, 
$   
J^{\pm } = \bar{N}\,\, (\alpha _{\pm S}^{(0)}+\alpha _{\pm S}^{(3)}\tau _{3}) \,N, 
$
where $N^T = (p, n)$ is the nucleon isospin doublet.
The isosclar $\alpha^{(0)}$ and the isovector $\alpha^{(3)}$ coefficients are 
\begin{eqnarray} \nn
\alpha _{V}^{(0)} &=&\frac{1}{8}(G_{V}^{u}+
G_{V}^{d})(\eta^{u1}+ \eta^{d1}),  \ \ \
\alpha _{V}^{(3)} = \frac{1}{8}%
(G_{V}^{u}-G_{V}^{d})(\eta^{u1}-\eta^{d1}),  \nonumber \\ \nn
\alpha_{\pm S}^{(0)} &=&\frac{1}{16}(G_{S}^{u}+G_{S}^{d})
(\eta_{L}^{d1}\pm \eta_{R}^{d1}) +
\frac{1}{8}G_S^{s}(\eta_{L}^{d2} \pm \eta_{R}^{d2}) +  
\frac{1}{8}G_S^{b}(\eta_{L}^{d3} \pm \eta_{R}^{d3}),\\
\alpha_{\pm S}^{(3)} &=& -\frac{1}{16}
(G_{S}^{u}-G_{S}^{d})(\eta_{L}^{d1}\pm \eta_{R}^{d1}),
\end{eqnarray}

Following the approach of Refs. \cite{FKKV:2000,KKS:2001} we derive from the Lagrangian
(\ref{nucl1}) the coherent $\mu-e$ conversion branching ratio in the form
\begin{equation}
R_{\mu e^-} = 
\frac{G^2_F }
{2 \pi } \ {\cal Q}\ \frac{p_e E_e({\cal M}_p + {\cal M}_n)^2}{\Gamma(\mu^-\to capture)} \, ,
\label{Rme}
\end{equation}
where $\Gamma(\mu^-\to capture)$ is the total rate of ordinary muon capture and 
\begin{eqnarray}
{\cal Q} \,=\, 2|\alpha_V^{(0)}|^2 +
|\alpha_{+S}^{(0)}|^2 + |\alpha_{-S}^{(0)}|^2 
+2\ {\rm Re}\{\alpha_V^{(0)}[\alpha_{+S}^{(0)}+\alpha_{-S}^{(0)}] \}\, .
\label{Rme.1}
\end{eqnarray}
In this formula we neglected the contribution of isovector currents which is small 
for most of the experimentally interesting nuclei \cite{FKKV:2000,KKS:2001}.
The numerical values of the nuclear matrix elements ${{\cal M}}_{p,n}$ 
for the currently interesting have been calculated in Ref. \cite{KKS:2001}. 
 
The most stringent experimental bounds on the branching ratio $R_{\mu e}$
have been set by the SINDRUM2 experiment (PSI) with $^{197}$Au
and $^{48}$Ti stopping targets:
\begin{eqnarray}\label{Au}
R_{\mu e}^{Au} &=& 
\frac{\Gamma(\mu^- + {}^{197}Au\rightarrow e^- + {}^{197}Au)} 
{\Gamma(\mu^- + {}^{197}Au \rightarrow \nu_{\mu} + ^{197}Pt)} 
\leq 5.0\times 10^{-13}\ , \ \ \ \mbox{(90\% C.L.)} \ \ \  \cite{Schaaf},\\
\label{Ti}
R_{\mu e}^{Ti} &=& 
\frac{\Gamma(\mu^- + \ \ {}^{48}Ti\rightarrow e^- + \ \ {}^{48}Ti)} 
{\Gamma(\mu^- + {}^{48}Ti \rightarrow \nu_{\mu} + \ \  ^{48}Sc)} 
\leq 6.1\times 10^{-13}\ , \ \ \ 
\mbox{(90\% C.L.)} \ \ \  \cite{SINDRUM}.
\end{eqnarray}
Note that a $^{48}$Ti target will be used in the future experiment planned at 
the muon factory at KEK (Japan) \cite{Aoki}. This experiment is going to 
increase the sensitivity up to $R_{\mu e} \le 10^{-18}$.

In the near future new bounds are expected from the MECO (Brookhaven) 
experiment with an $^{27}$Al target  
\begin{eqnarray}\label{Al}
R_{\mu e}^{Al}  = 
\frac{\Gamma(\mu^- + {}^{27}Al\rightarrow e^- +  {}^{27}Al)}
{\Gamma(\mu^- + {}^{27}Al \rightarrow \nu_{\mu} +  {}^{27}Mg)} 
\leq 2\times 10^{-17}\ \ \ \ \ \ \ \   \cite{MECO}
\end{eqnarray}

From these experimental limits it is straightforward to extract upper bounds 
on various products of the type $\lambda\lambda'$, $\lambda'\lambda'$. 
Many of them have been previously derived in Refs. 
\cite{9701381}-\cite{FKKV:2000}. 
In Table 1 we present the new upper bounds that are associated with the b-quark contribution.
We show the three cases, corresponding to the experimental limits in Eqs. (\ref{Au})-(\ref{Al}).
%
\begin{table}[h!]
\begin{center}
\begin{tabular}{|c|c|c|c|c|}
\hline \hline
Parameters & Previous limits & Present limits (Au)& 
Present limits (Ti) & Expected limits (Al) \\
& & $^{197}$Au$(\mu-e)\cdot B_1$ & $^{48}$Ti$(\mu-e)\cdot B_2$  & 
$^{27}$Al$(\mu-e)\cdot B_3$ \\
\hline
& & & &  \\
$|\lambda '_{133}\, \lambda_{121}|$ &$6.9\cdot 10^{-5}$ & $8.5\cdot 10^{-7}$   &
$2.0\cdot 10^{-6}$&  $1.2\cdot 10^{-8}$\\
$|\lambda '_{233}\, \lambda_{212}|$ &$7.4\cdot 10^{-3}$ & $8.5\cdot 10^{-7}$  &
$2.0\cdot 10^{-6}$&  $1.2\cdot 10^{-8}$\\
$|\lambda '_{333}\, \lambda_{312}|$ &$2.8\cdot 10^{-2}$ & $8.5\cdot 10^{-7}$   &
$2.0\cdot 10^{-6}$&  $1.2\cdot 10^{-8}$\\
$|\lambda '_{333}\, \lambda_{321}|$ &$3.2\cdot 10^{-2}$ & $8.5\cdot 10^{-7}$  &
$2.0\cdot 10^{-6}$&  $1.2\cdot 10^{-8}$\\
& & & &  \\
\hline
\hline
\end{tabular}
\caption{The new upper bounds from the b-quark contribution to \m conversion.
The previous bounds were taken from \protect\cite{Barb}. 
The scaling factors $B_{1,2,3}$ are defined in the text.}
\end{center}
\end{table}
%

In the derivation of these bounds we assumed, as usual, the dominance of only one of these 
products with the specific combination of generation indices. We also assumed that all 
the scalar masses in Eq. (\ref{coeff12}) are equal 
$\tilde m_{{u}L(n)}\approx \tilde m_{{d}L,R(n)} \approx \tilde m_{\nu(n)}\approx \tilde m$.

As can be seen from Table 1, our new limits(columns 3-5) are significantly more stringent than 
those previously known in the literature \cite{Barb} (column 2).
In Table 1 the quantities $B_{1,2,3}$ denote scaling factors defined as 
\begin{eqnarray}\label{scale}
B_1 = (R^{exp}_{\mu e}/5.0 \cdot 10^{-13})^{1/2},\ \ \ 
B_2 = (R^{exp}_{\mu e}/6.1 \cdot 10^{-13})^{1/2},\ \ \
B_3 = (R^{exp}_{\mu e}/2.0\cdot 10^{-17})^{1/2}.
\end{eqnarray}
They allow recalculating limits given in Table 1 to the case corresponding 
to the experimental upper bounds on the branching ratio $R^{exp}_{\mu e}$ 
other than in Eqs. (\ref{Au})-(\ref{Al}).

In summary, we found new important contribution to $\mu^--e^-$ conversion
originating from the b-quark sea of the nucleon.
We have shown that among the heavy $c,b,t$-quarks 
only the b-quark can contribute to the coherent \m conversion via 
the scalar interactions involving down quarks $d,s,b$. The heavy quarks contribution
to the scalar current is materialized with the gluon exchange shown in the triangle 
diagram of Fig. 1. From the existing data and expected experimental constraints on 
the branching ratio $R_{\mu e^-}$ we obtained new upper limits on the products of 
the trilinear \r parameters of the type $\lambda_{n12}\lambda'_{n33}$, 
$\lambda_{n21}\lambda'_{n33}$   
which are significantly more stringent than those existing in the literature.

\bigskip

This work was supported in part by Fondecyt (Chile) under grant
8000017, by a C\'atedra Presidencial (Chile) and by RFBR (Russia) under 
grant 00-02-17587.

\end{document}